# A theoretical investigation on the carrier mobilities of armchair silicene nanoribbons

Guo Wang

*Department of Chemistry, Capital Normal University, Beijing 100048, China*

*E-mail address:* wangguo@mail.cnu.edu.cn

ABSTRACT

Armchair silicene nanoribbons with width of 9-39 silicon atoms are investigated by using self-consistent field crystal orbital method based on density functional theory. The carrier mobilities obtained from deformation potential theory oscillate with respect to the width and the values are a fraction of what the graphene nanoribbons have. The buckled structure, hydrogen saturation, edge reconstruction as well as edge roughness decrease the carrier mobilities which are explained with the aid of crystal orbitals.

## 1. Introduction

The scale of silicon-based electronics is already down to tens of nanometers with fast-developing modern computer industry. Graphene is one potential candidate of future electronic devices for its extremely high carrier mobility [1]. Silicene [2], a counterpart of graphene, was recently prepared on silver surfaces [3,4]. Silicene nanoribbons with width of about several nanometers were also fabricated on silver surfaces [5,6].

Silicene is a two-dimensional zero gap semiconductor [2,7]. However, a finite band gap is essential for room temperature on/off operations. One-dimensional armchair silicene nanoribbons are predicted to exhibit band gaps [7-11], just as graphene nanoribbons. Theoretical studies indicated that silicene (nanoribbons) have many similar electronic properties with those of graphene (nanoribbons) [5,6], such as magnetism [8,12,13] and half-metallic property under electric filed [9].

With the size of silicon-based electronic devices going down, one atomic thick silicon layer will come into the scope of computer industry. Silicene nanoribbons could be a candidate of next generation electronic devices before carbon-based electronic devices coming to large-scale application. Here, an important question is whether silicene nanoribbons have high carrier mobilities which are crucial to the high-performance electronic devices.

In graphene nanoribbons, edge state is an important factor affecting their carrier mobilities [14-16]. This should be also important to silicene nanoribbons. Furthermore, silicene has low-buckled structure [2,8,9,17,18] with $sp^2/sp^3$ mixed hybridization rather than planar structure, which will also affect their carrier mobilities. It is interesting to explore these types of influence before the application of silicene nanoribbons in electronic devices.

In the present work, carrier mobilities of silicene nanoribbons with width of 9-39 silicon atoms are studied based on density functional theory. The influence coming from the buckled geometry and different edge states on the carrier mobilities is investigated with the aid of crystal orbital analysis.

## 2. Theory and computational details

The carrier mobilities of the armchair silicene nanoribbons can be obtained from deformation potential theory. In the theory, the energy of acoustic phonons is assumed to be comparable to the carrier thermal energy, and carriers are scattered mostly by acoustic phonons [19]. The carrier mobility for one-dimensional system is obtained by

$$\mu = \frac{e\hbar^2 C}{(2\pi k_B T)^{1/2} |m^*|^{3/2} E_1^2}, \qquad (1)$$

where carrier effective mass $m^*$ is obtained by fitting frontier band structures, stretching modulus $C$ as well as deformation potential constant $E_1$ in longitudinal acoustic phonon scattering mode are obtained under deformed geometries. The deformation potential approach has been successfully applied to similar systems, such as conducting polymers [20], graphene nanoribbons [21-23] and carbon nanotubes [24,25].

Armchair silicene nanoribbons with width of 9-39 silicon atoms are investigated, which are denoted as $N$-ASiNR ($N$=9-39). The edges of the ASiNRs are passivated with hydrogen atoms. Furthermore, structures with planar geometry and with different edge states are also calculated for comparison. Full optimization of the nanoribbons is performed under Bloch functions constructed with standard 6-21G(*d*, *p*) basis set which are included in the self-consistent field crystal orbital program CRYSTAL06 [26]. Since the carrier mobility depends on powers of the parameters, the sophisticated hybrid density functional B3LYP [27] (VWN5 [28]), which uses Becke's three parameters [29] to incorporate a portion of exact exchange and LYP correlation functional [30], is adopted to evaluate the energy and related properties accurately.

Atomic grid with 55 radial and 434 angular points is used in order to evaluate the density functional numerical integrations. Default values of convergence criteria in the CRYSTAL06 program are used. A Monkhorst-Pack sampling with 41 *k*-points in the first Brillouin zone is sufficient to obtain the converged structure, energy and derived electronic properties. When calculating the band structures, a uniform *k*-points sampling with 801 points is used in order to facilitate the fitting of carrier effective mass at a certain range. The thermal energy of carriers is usually represented by $k_B T$. However, without a very sharp density of state near edge of frontier bands, carriers in a wider energy range could participate in the real conduction process of devices. In this work, $10 k_B T$ [31] is attempted to obtain the carrier effective mass and carrier mobility.

## 3. Results and discussion

### 3.1 Band gaps

The optimized ASiNRs have low-buckled geometry. For example, the lengths of Si-Si bonds at the center are both 0.228 nm for the narrowest 9-ASiNR and the widest 39-ASiNR. The POAV angles [32] are 100.96 and 100.95°, respectively. These values are very close to the 0.228 nm and 101.73° for two-dimensional silicene obtained from plane wave calculations [18]. It is indicated that besides silicene, ASiNRs are also composed of silicon atoms with $sp^2/sp^3$ mixed hybridization. In the low-buckled geometry, there is also $\pi$ character in the Si-Si bonds, so there are considerable dispersions in the frontier bands.

Since the band structures are similar for all these ASiNRs, only band structures of the widest 39-ASiNR are shown in the inset of Figure 1(a). Direct band gaps exist at the center of the first Brillouin zone between the two frontier bands for all the ASiNRs.

From Figure 1(a), it can be seen that band gap oscillation with respect to $N$ exists for all these ASiNRs. The ASiNRs falls into three families: $N=3q$, $N=3q+1$ and $N=3q+2$, where $q$ is a positive integer. Guided by the dotted lines in Figure 1(a), the band gap sequence is $\text{gap}_{3q+1} > \text{gap}_{3q} > \text{gap}_{3q+2}$. It can be attributed to different quantum confinements for different widths of the ASiNRs when two-dimensional silicene reduces to one-dimensional structures. The sequence agrees well with plane wave calculated results [8-10] and it is also agrees with that of graphene nanoribbons [23,33] because of similar valence electronic structures between carbon and silicon atoms. The band gaps of the ASiNRs shown in Figure 1(a) are all less than 1 eV. For the ASiNRs with $N=3q+2$, the band gaps are less than 0.1 eV. The band gaps are smaller than those of the graphene nanoribbons obtained with the same method [23]. However, it is similar to graphene nanoribbons that the band gap decreases with respect to the width $N$ and will reach zero for the two-dimensional structure.

*3.2 Stretching moduli*

The stretching moduli of 9-ASiNR and 39-ASiNR listed in Table 1 are 1033 and 4339 eVnm$^{-1}$, respectively. From 9-ASiNR to 39-ASiNR, the stretching modulus increases almost linearly with the width of silicon atoms. There is an average 109 eVnm$^{-1}$ increment of stretching modulus when the ASiNR is one atom wider, which is only 31% magnitude of that for graphene nanoribbons (350 eVnm$^{-1}$) [23]. It is the longer $\pi$ bond between the two 3$p$ orbitals of silicon atoms that makes the bond weaker than that in graphene nanoribbons.

Considering the calculated average ASiNR width of 0.194 nm per silicon atom and van der Waals distance of 0.420 nm for silicon [34], the elastic constant $C/A_0$ ($A_0$ is the cross section along the one-dimensional direction) is obtained averagely as 0.214

TPa. This value is far less than the extremely high value 1.34 TPa for graphene nanoribbons obtained with the same method [23], due to the longer bond lengths and weaker bonds.

### 3.3 Deformation potential constants

The deformation potential constants for the two types of carrier are different. For 39-ASiNR, the valence ($E_{1v}$) or conduction ($E_{1c}$) band deformation potential constant listed in Table 1 are 4.63 and 2.88, respectively. Actually $E_{1v}$ is larger than $E_{1c}$ for all the ASiNRs with $N=3q$, while $E_{1c}$ is larger for $N=3q+1$ and $N=3q+2$. The "large" ("small") deformation potential constants are in the range of 4.25-5.06 (2.45-3.43) eV. It is noted that the deformation potential constants are smaller than those of the graphene nanoribbons (10.1-13.9 or 3.1-5.7 eV for the "large" or "small" constants), due to the weak bonds in ASiNRs.

Compared with the mechanical property, there is more quantum effect in the deformation potential. When the crystal orbital is delocalized with respect to the one-dimensional deformation direction, the energy change in the deformation process should be smaller than that for the localized crystal orbital [21-23,25], because the consequence could be partly compensated by large conjugated system. For 39-ASiNR, the highest occupied crystal orbital (HOCO) shown in Figure 2(a) is localized orbital with respect to the one-dimensional direction, while the lowest unoccupied crystal orbital (LUCO) in Figure 2(b) is delocalized. This is why $E_{1v}$ is larger than $E_{1c}$ for 39-ASiNR. The difference of valence and conduction band deformation potential constants makes the mobility of one type of carrier larger than another's, which is the interesting carrier polarity [21] phenomenon also existing in graphene nanoribbons [21-23] and carbon nanotubes [24,25].

### 3.4 Carrier effective masses

The hole and electron effective masses fitted within 10 $k_B T$ ($T$ = 298.15 K, room temperature) from the frontier orbitals are in the range of 0.13-0.91 and 0.13-1.16 $m_0$ ($m_0$ is the mass of a free electron) for the ASiNRs, which are shown in Figure 1(b) and 1(c). The carrier mass of bulk silicon [35] also falls in this range. Carriers in the ASiNRs are much heavier than in the graphene nanoribbons (0.070-0.206 and 0.071-0.222 $m_0$ for holes and electrons [23]) due to the weaker conjugated $\pi$ bonds in ASiNRs. For $N=3q$ and $3q+1$, carrier mass varies little with the width and is in the range of 0.1-0.2 $m_0$. However, for $N=3q+2$ indicated by dotted lines in Figure 1(b) and 1(c), carrier mass is much larger and increases with the width notably. For 38-ASiNR, carrier masses are around a free electron and are no longer as small as those of the other two families.

One possible explanation could be that when $N=3q+2$, the band gaps of the ASiNRs are all less than 0.1 eV and the band gap decreases with the width. With such a near zero band gap, an ASiNR especially for 38-ASiNR has some "metallic" properties. Electrons in many metals act as nearly free electrons. [36] Their effective masses are around an $m_0$ and much larger than 0.2 $m_0$ for ASiNRs with $N=3q$ and $3q+1$. Carriers in the two very near frontier bands of the ASiNRs with $N=3q+2$ have much larger effective masses.

*3.5 Carrier mobilities*

The carrier mobilities of the ASiNRs oscillate with the width $N$, which are shown in Figure 1(d). For $N=3q+2$ indicated by the dotted line, the room temperature hole and electron mobilities are all less than 1000 cm$^2$V$^{-1}$s$^{-1}$. These values are much smaller than those of the other two families, because of the bigger carrier masses. For $N=3q$ and $3q+1$, several mobilities are above 5000 cm$^2$V$^{-1}$s$^{-1}$.

As to the polarity, electron mobility is larger than hole mobility for $N=3q$, while hole mobility is larger for $N=3q+1$ or $3q+2$. This is mainly caused by the different orientation of frontier crystal orbitals and then the unbalanced deformation potential constants for the two types of carriers. For example, $E_{1c}$ is smaller than $E_{1v}$ for 39-ASiNR and 9-ASiNR listed in Table 1, so the electron mobilities are larger than the hole mobilities according to equation (1).

ASiNRs have smaller stretching moduli than graphene nanoribbons due to the weak $\pi$ bonds between $3p$ orbitals. Furthermore, carriers are heavier in ASiNRs than in graphene nanoribbons. These should make the carrier mobilities much smaller than those of the graphene nanoribbons. However, the weak $\pi$ bonds in ASiNRs also result in small deformation potential constants, which compensate part of the carrier mobility decrement. Finally, the major carrier mobilities are a fraction of what the graphene nanoribbons have [23]. For example, the electron mobility of 39-ASiNR (5965 $cm^2V^{-1}s^{-1}$) is 22% magnitude of the graphene nanoribbons with the same number of atoms. The proportions are 15%-44%, 28%-92% and 4%-20% for $N=3q$, $3q+1$ and $3q+2$, respectively. The lowest proportion for $N=3q+2$ is due mainly to the near zero band gaps and the big carrier masses. For other two groups, they are still quite high for modern silicon industry.

Normally, a suitable band gap is necessary for device operation. For ASiNRs with $N=3q+2$, the band gaps are too small to give good on/off ratio. Fortunately, these ASiNRs have low carrier mobilities. Width control of ASiNRs is beneficial to the application and might be achieved in future silicon industry.

*3.6 Temperature effect*

The widest 39-ASiNR is taken as an example to explore the temperature effect. Carrier masses fitted within different energy ranges from $k_BT$ to $10k_BT$ ($T$=298.15 K) are shown in Figure 1(e). Carriers are lighter near the center of the Brillouin zone, which is similar to the graphene nanoribbons [23] and also the fact that the carrier mass in graphene varies with the probed energy in experiments [37,38]. In Figure 1(e), there is no large difference between hole and electron masses. From $10k_BT$ to $k_BT$, electron masses decrease from 0.169 to 0.027 $m_0$. This should contribute 14.7 times increment of electron mobility according to equation (1). Furthermore narrower energy range corresponds to lower temperature. Considering that mobility is also related to $T^{-1/2}$ in equation (1), the electron mobility at 29.815 K is 299344 $cm^2V^{-1}s^{-1}$, which is 50 times as large as that at room temperature (5965 $cm^2V^{-1}s^{-1}$). The temperature dependence of carrier mobility is also shown in Figure 1(e). Because carriers are lighter near the frontier band edges, the carrier mobility dependence should be stronger than $T^{-1/2}$.

Usually the overheated chips have low performance. At 373.15 K (100 ℃), the hole (electron) mass of 39-ASiNR increases from 0.155 (0.169) $m_0$ to 0.326 (0.420) $m_0$, while the hole (electron) mobility decreases from 2643 (5965) cm$^2$V$^{-1}$s$^{-1}$ to 776 (1362) cm$^2$V$^{-1}$s$^{-1}$. The performance at 100 ℃ is only 20-30% as good as those at room temperature.

*3.7 Other structures*

Unlike graphene nanoribbons, the most stable ASiNRs have low-buckled structures. However, planar structure of 39-ASiNR is also calculated for comparison. In Figure 2(c) and 2(d), the frontier crystal orbitals of 39-ASiNR(planar) are similar with those of 39-ASiNR, but the crystal orbitals are more uniform and the $\pi$ bonds conjugation is better for 39-ASiNR(planar). This results in slightly lighter masses and higher deformation potential constants. The planar structure with $sp^2$ hybridization also has higher stretching modulus. The increment of hole (electron) mobility listed in Table 1 and Figure 1(f) is 27% (17%).

Edges of graphene nanoribbons could be saturated upon hydrogenation [39]. The structure with $sp^3$ hybridized edges is also investigated. Each edge silicon atom is attached with two hydrogen atoms. The optimization indicates that low-buckled geometry of 39-ASiNR(2H) shown in Figure 2(e) also exists at the non-edge silicon atoms. In Figure 2(e) and 2(f), there is no orbital on the saturated silicon atoms at the edges. The carrier masses listed in Table 1 increase a little. The polarity changes and the hole mobility is larger than the electron one, because the number of silicon atoms participating in the conjugated system has changed and the $\pi$ system of 39-ASiNR(2H) should belong to $3q+1$ family after excluding the two edges. The hole mobility of 39-ASiNR(2H) in Table 1 is 93% as large as the electron mobility of 39-ASiNR due to the reduced conjugated system.

A silicane nanoribbon is formed when all silicon atoms are transferred to $sp^3$ hybridized atoms upon hydrogenation [6]. Being different from other band structures in this work, the direct band gap of 39-ASilicaneNR occurs at the edge of the first Brillouin zone. The frontier crystal orbitals shown in Figure 2(g) and 2(h) are mainly edge states. The conjugated system becomes much smaller which results in much heavier carriers (24.45 $m_0$ for holes in Table 1). The mobilities of the confined carriers are significantly reduced especially for holes (3 cm$^2$V$^{-1}$s$^{-1}$ in Table 1) due to the narrowest frontier crystal orbital in Figure 2(g).

When there is no hydrogen termination at the edges, an ASiNR could undergo a reconstruction to form a self-passivated structure. The calculated reconstructed mode of 39-ASiNR(reconstruction) shown in Figure 2(i) is consistent with the plane wave calculated result [8]. The frontier crystal orbitals shown in Figure 2(i) and 2(j) are similar with those of 39-ASiNR except that the conjugation at the edges is not as good as that in the middle. This makes a little heavier carrier masses which decrease the mobilities. Smaller deformation potential constants caused by weaker bonds cancel out part of this effect. The final carrier mobility in Table 1 has only 13% or 7% decrement for holes and electrons compared with those of 39-ASiNR.

Edge roughness is an important issue affecting the performance of graphene nanoribbon devices [14-16]. The rough structure constructed from the narrowest 9-ASiNR is calculated in order to reduce large computational cost. The frontier crystal orbitals of 9-ASiNR(roughness) shown in Figure 2(k) and 2(l) indicate that $\pi$ conjugation also exists in the rough structure. However, the conjugation at the edges needs to extend along the atomic lines. Some of the orbitals are missing when there is no atom at the rough edges. The weaker conjugation gives heavier carrier masses and smaller stretching moduli. Some crystal orbitals along the rough edges increase the proportion of localized bonds with respect to the one-dimensional direction and result in larger deformation potential constants. The hole or electron carrier mobility in Table 1 reduces to 51% or 34% magnitude of 9-ASiNR.

## 4. Conclusions

Armchair silicene nanoribbons with width of 9-39 silicon atoms are investigated by using self-consistent field crystal orbital method based on density functional theory. It is indicated that band gap oscillation occurs for these nanoribbons, depending on their width of silicon atoms $N$. The carrier mobilities obtained under deformation potential theory are in the order of $10^3$ cm$^2$V$^{-1}$s$^{-1}$ except for $N=3q+2$ due to the narrow band gaps and big carrier masses. The major carrier mobilities for $N=3q$ and $N=3q+1$ are 15%-44% and 28%-92% magnitude of what graphene nanoribbons have. Some of the mobilities are above 5000 cm$^2$V$^{-1}$s$^{-1}$ which are still large for modern silicon-based electronics.

The carrier masses are lighter near the band edges, so the temperature dependence of the carrier mobility is stronger than $T^{-1/2}$. As shown in Figure 1(f), planar structure 39-ASiNR(planar) is beneficial to the $\pi$ bonds conjugation and then the carrier mobility. Silicane nanoribbons do not have high carrier mobilities due to lacking of large conjugated systems. Edge-hydrogen saturation, edge reconstruction and edge roughness decrease the carrier mobilities to some extend, but the major carrier mobilities are comparable with the corresponding structures without defects, for they still have large $\pi$ conjugated systems. With the size of fast electronic devices going down, single-layer armchair silicene nanoribbons could be the next candidate with high carrier mobilities before graphene nanoribbons, because of their compatibility with current silicon industry.

**Table 1.** The stretching modulus $C$ (eVnm$^{-1}$), valence band deformation potential constant $E_{1v}$, conduction band deformation potential constant $E_{1c}$ (eV), hole effective mass $|m_h^*|$, electron effective mass $|m_e^*|$ ($m_0$), hole mobility $\mu_h$, and electron mobility $\mu_e$ (cm$^2$V$^{-1}$s$^{-1}$) of the nanoribbons.

|  | $C$ | $E_{1v}$ | $E_{1c}$ | $|m_h^*|$ | $|m_e^*|$ | $\mu_h$ | $\mu_e$ |
|---|---|---|---|---|---|---|---|
| 39-ASiNR | 4339 | 4.63 | 2.88 | 0.16 | 0.17 | 2643 | 5965 |
| 39-ASiNR(planar) | 4879 | 4.79 | 3.54 | 0.14 | 0.13 | 3365 | 7000 |
| 39-ASiNR(2H) | 4370 | 3.05 | 4.41 | 0.17 | 0.18 | 5543 | 2315 |
| 39-ASilicaneNR | 3779 | 2.72 | 2.96 | 24.45 | 0.90 | 3 | 404 |
| 39-ASiNR(reconstruction) | 4370 | 4.41 | 2.45 | 0.18 | 0.22 | 2291 | 5558 |
| 9-ASiNR | 1033 | 4.25 | 2.45 | 0.14 | 0.13 | 903 | 2772 |
| 9-ASiNR(roughness) | 957 | 4.73 | 3.27 | 0.18 | 0.18 | 464 | 943 |

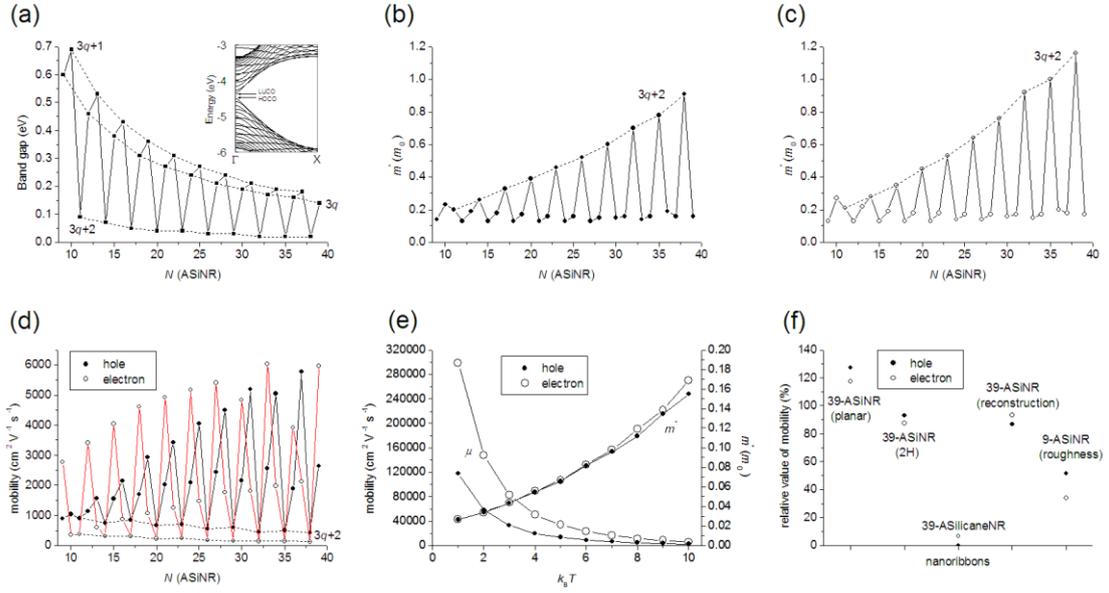

**Figure 1.** (a) Band gaps of *N*-ASiNRs (inset: band structures of 39-ASiNR), (b) hole and (c) electron effective masses of *N*-ASiNRs, (d) carrier mobilities of *N*-ASiNRs, (e) temperature effect for 39-ASiNR, (f) relative carrier mobilities of the nanoribbons with respect to the corresponding *N*-ASiNR.

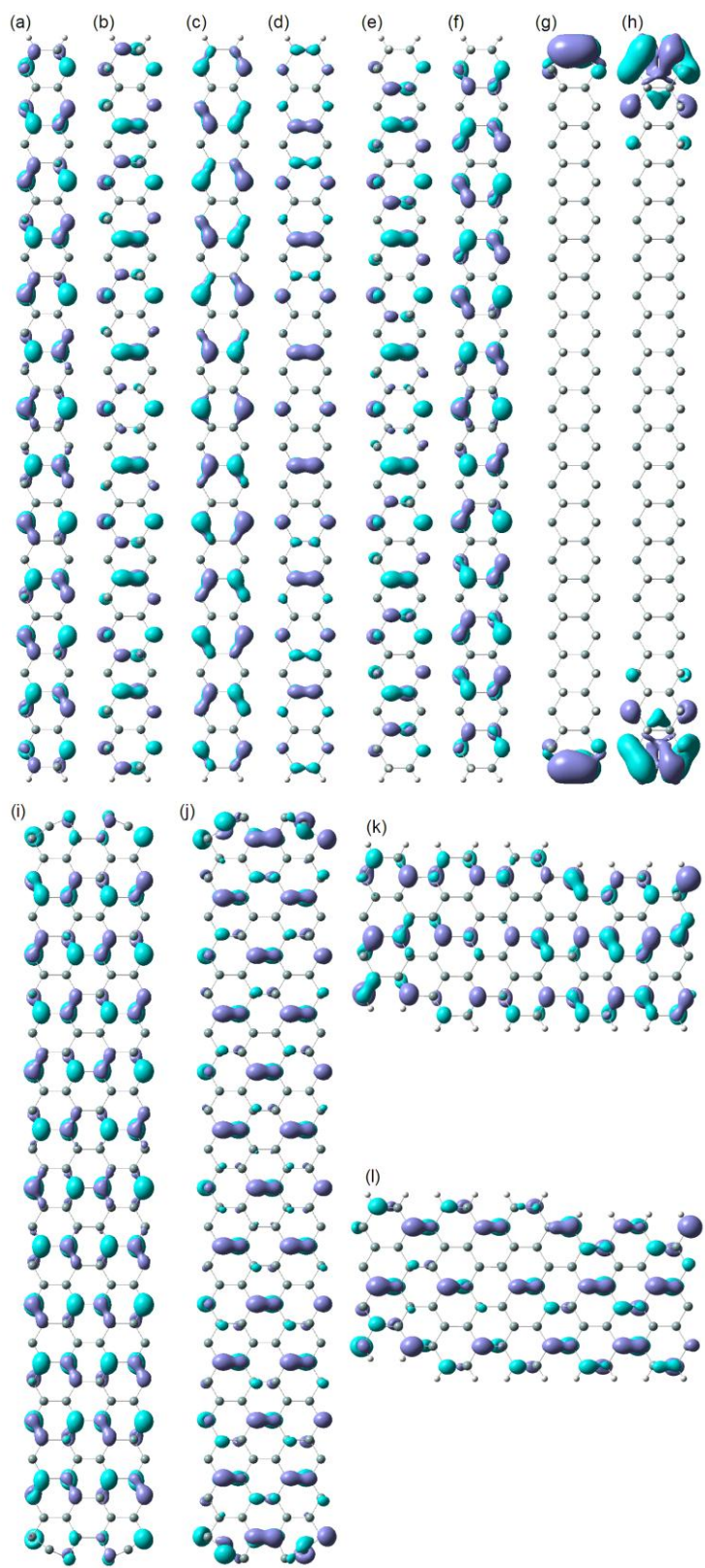

**Figure 2.** Frontier crystal orbitals of the nanoribbons. (a) HOCO and (b) LUCO of 39-ASiNR, (c) HOCO and (d) LUCO of 39-ASiNR(planar), (e) HOCO and (f) LUCO of 39-ASiNR(2H), (g) HOCO and (h) LUCO of 39-ASilicaneNR, (i) HOCO and (j) LUCO of 39-ASiNR(reconstruction), (k) HOCO and (l) LUCO of 9-ASiNR(roughness).